\documentclass[twocolumn,showpacs,prl]{revtex4-1}
\usepackage{graphicx}
\usepackage{color}
\usepackage{amssymb}

\newcommand{\xscrpt}[1]{{\textstyle \mbox{\scriptsize #1}}}

\newcommand{\ntsj}[2]{#1\,{}^{3\!}S_{#2}}
\newcommand{\ntdj}[2]{#1\,{}^{3\!}D_{#2}}
\newcommand{\tsj}[1]{{}^{3\!}S_{#1}}
\newcommand{\tdj}[1]{{}^{3\!}D_{#1}}
\begin{document}
\title{Observation of the $\Upsilon(\ntdj{2}{1})$ and
indication of the $\Upsilon(\ntdj{1}{1})$}
\author{Eef van Beveren}
\email{eef@teor.fis.uc.pt}
\affiliation{Centro de F\'{\i}sica Computacional,
Departamento de F\'{\i}sica, Universidade de Coimbra,
P-3004-516 Coimbra, Portugal}
\author{George Rupp}
\email{george@ist.utl.pt}
\affiliation{Centro de F\'{\i}sica das Interac\c{c}\~{o}es Fundamentais,
Instituto Superior T\'{e}cnico, Edif\'{\i}cio Ci\^{e}ncia, Piso 3,
P-1049-001 Lisboa, Portugal}
\date{\today}

\begin{abstract}
We present evidence, from published BABAR data, for
the existence of further excitations
in the $b\bar{b}$ spectrum,
namely the long-awaited $\Upsilon(\ntdj{1}{1})$ and $\Upsilon(\ntdj{2}{1})$,
with central masses
of 10098$\pm$5 MeV and 10495$\pm$5 MeV, respectively
The significance of the $\Upsilon(\ntdj{1}{1})$ and $\Upsilon(\ntdj{2}{1})$
signals is found to be 3.0 and 10.7 standard deviations, respectively.
\end{abstract}


\maketitle

In Ref.~\cite{PRD78p112002} the BABAR Collaboration
presented an analysis of data on
$e^{+}e^{-}$ $\to$ $\pi^{+}\pi^{-}\Upsilon (2S,1S)$
$\to$ $\pi^{+}\pi^{-}e^{+}e^{-}$
(and $\pi^{+}\pi^{-}\mu^{+}\mu^{-}$),
with the aim to study hadronic transitions between
$b\bar{b}$ excitations and the $\Upsilon (1S)$ and $\Upsilon (2S)$,
based on 347.5 fb$^{-1}$ of data
taken with the BABAR detector at the PEP-II storage rings.
BABAR reported \cite{PRD78p112002} the first observation
of the $\Upsilon (10580)\to\eta\Upsilon (1S)$ decay,
with a branching fraction of
${\cal B}(\Upsilon (10580)\to\eta\Upsilon (1S))$
$=$ $(1.96\pm 0.06_\xscrpt{stat}\pm 0.09_\xscrpt{syst})\times 10^{-4}$,
measured the value $2.41\pm 0.40_\xscrpt{stat}\pm 0.12_\xscrpt{syst}$
for the ratio of partial widths
$\Gamma\left[\Upsilon (10580)\right.$ $\to$
$\left.\eta\Upsilon (1S)\right]$ $/$
$\Gamma\left[\Upsilon (10580)\right.$ $\to$
$\left.\pi\pi\Upsilon (1S)\right]$,
set 90\%-confidence-level upper limits on the ratios
$\Gamma\left[\Upsilon (2S)\right.$ $\to$
$\left.\eta\Upsilon (1S)\right]$ $/$
$\Gamma\left[\Upsilon (2S)\right.$ $\to$
$\left.\pi\pi\Upsilon (1S)\right]$
$<5.2\times 10^{-3}$ and
$\Gamma\left[\Upsilon (3S)\right.$ $\to$
$\left.\eta\Upsilon (1S)\right]$ $/$
$\Gamma\left[\Upsilon (3S)\right.$ $\to$
$\left.\pi\pi\Upsilon (1S)\right]$
$< 1.9\times 10^{-2}$
and finally
presented new measurements of the ratios
$\Gamma\left[\Upsilon (10580)\right.$ $\to$
$\left.\pi\pi\Upsilon (2S)\right]$ $/$
$\Gamma\left[\Upsilon (10580)\right.$ $\to$
$\left.\pi\pi\Upsilon (1S)\right]
=1.16\pm 0.16_\xscrpt{stat}\pm 0.14_\xscrpt{syst}$ and
$\Gamma\left[\Upsilon (3S)\right.$ $\to$
$\left.\pi\pi\Upsilon (2S)\right]$ $/$
$\Gamma\left[\Upsilon (3S)\right.$ $\to$
$\left.\pi\pi\Upsilon (1S)\right]$
$=$ $0.577\pm 0.026_\xscrpt{stat}\pm 0.060_\xscrpt{syst}$.

In the present paper the data of Ref.~\cite{PRD78p112002}
are further analyzed
for the possible existence
of the $\Upsilon (\ntdj{1}{1})$ and $\Upsilon(\ntdj{2}{1})$ states,
which we expect at roughly
$2\times 4.724+0.19\times(2n+2+1.5)=10.113$
and 10.493 GeV \cite{PRD27p1527}, for $n=0$ and
$n=1$, respectively.

BABAR published \cite{PRD78p112002}
the invariant-mass distributions of 34513 events for the reactions
$e^{+}e^{-}$ $\to$ $\pi^{+}\pi^{-}\Upsilon (2S,1S)$
$\to$ $\pi^{+}\pi^{-}\mu^{+}\mu^{-}$,
and those of 53627 events for the reactions
$e^{+}e^{-}$ $\to$ $\pi^{+}\pi^{-}\Upsilon (2S,1S)$
$\to$ $\pi^{+}\pi^{-}e^{+}e^{-}$.
The events were preselected for the transitions
$e^{+}e^{-}$ $\to$ $\pi^{+}\pi^{-}\Upsilon (1S)$
and
$e^{+}e^{-}$ $\to$ $\pi^{+}\pi^{-}\Upsilon (2S)$,
according to the invariant masses
of the final-state $e^{+}e^{-}$ and $\mu^{+}\mu^{-}$ pairs.
In the analysis, BABAR chose a rather large energy window
around the $\Upsilon (1S)$ and $\Upsilon (2S)$ masses,
of the order of $\pm$250 MeV,
for the invariant mass
of the final-state $e^{+}e^{-}$ and $\mu^{+}\mu^{-}$ pairs,
but imposed further cuts on the sample of published data.
In the present work, we shall apply much smaller windows,
viz.\ $\pm$20--30 MeV, though without imposing more cuts.

\begin{figure}[htbp]
\begin{center}
\begin{tabular}{c}
\includegraphics[height=170pt]{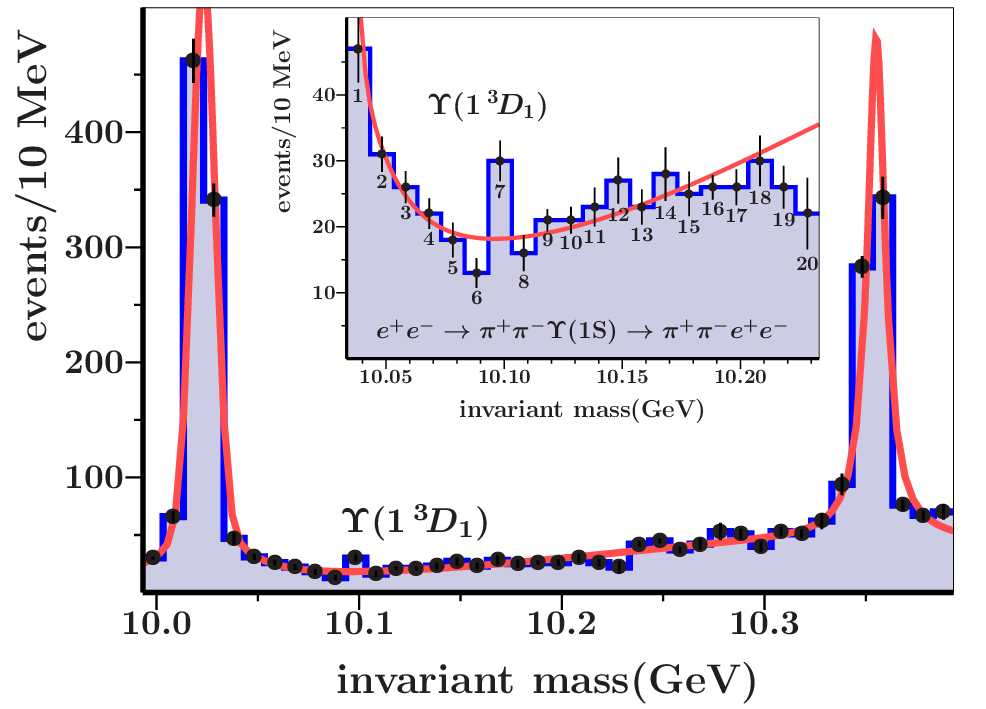}\\ [-10pt]
\end{tabular}
\end{center}
\caption{\small
Invariant-mass distribution for
2809 events (501 in the inset) in
$e^{+}e^{-}$ $\to$ $\pi^{+}\pi^{-}\Upsilon (1S)$
$\to$ $\pi^{+}\pi^{-}e^{+}e^{-}$.
The data are taken from Ref.~\cite{PRD78p112002}.
A window of $\Delta =30$ MeV around the $\Upsilon (1S)$
is taken for the allowed invariant masses of the $e^{+}e^{-}$ pair.
The bin width is 10 MeV.
We have indicated our fit to the background by a solid curve.
The $\Upsilon(\ntdj{1}{1})$ signal
has a relevance of 3.0$\sigma$.}
\label{1Ddata}
\end{figure}

For a thorough analysis of the data we shall restrict
ourselves to events from the BABAR sample of
$e^{+}e^{-}$ $\to$ $\pi^{+}\pi^{-}\Upsilon (1S)$
$\to$ $\pi^{+}\pi^{-}e^{+}e^{-}$ data.
In Fig.~\ref{1Ddata} we show
the invariant-mass distribution of the sample,
for a window size of $\pm\Delta$ (with $\Delta =30$ MeV)
around the $\Upsilon (1S)$ mass,
and for a data binning of 10 MeV.
Besides the dominant peaks of
the $\Upsilon (2S)$ and $\Upsilon (3S)$,
we observe a small enhancement exactly
where the data trend has a minimum due to
the large signals.
The enhancement is well visible
in the inset of Fig.~\ref{1Ddata}.

\begin{figure}[htbp]
\begin{center}
\begin{tabular}{c}
\includegraphics[height=320pt]{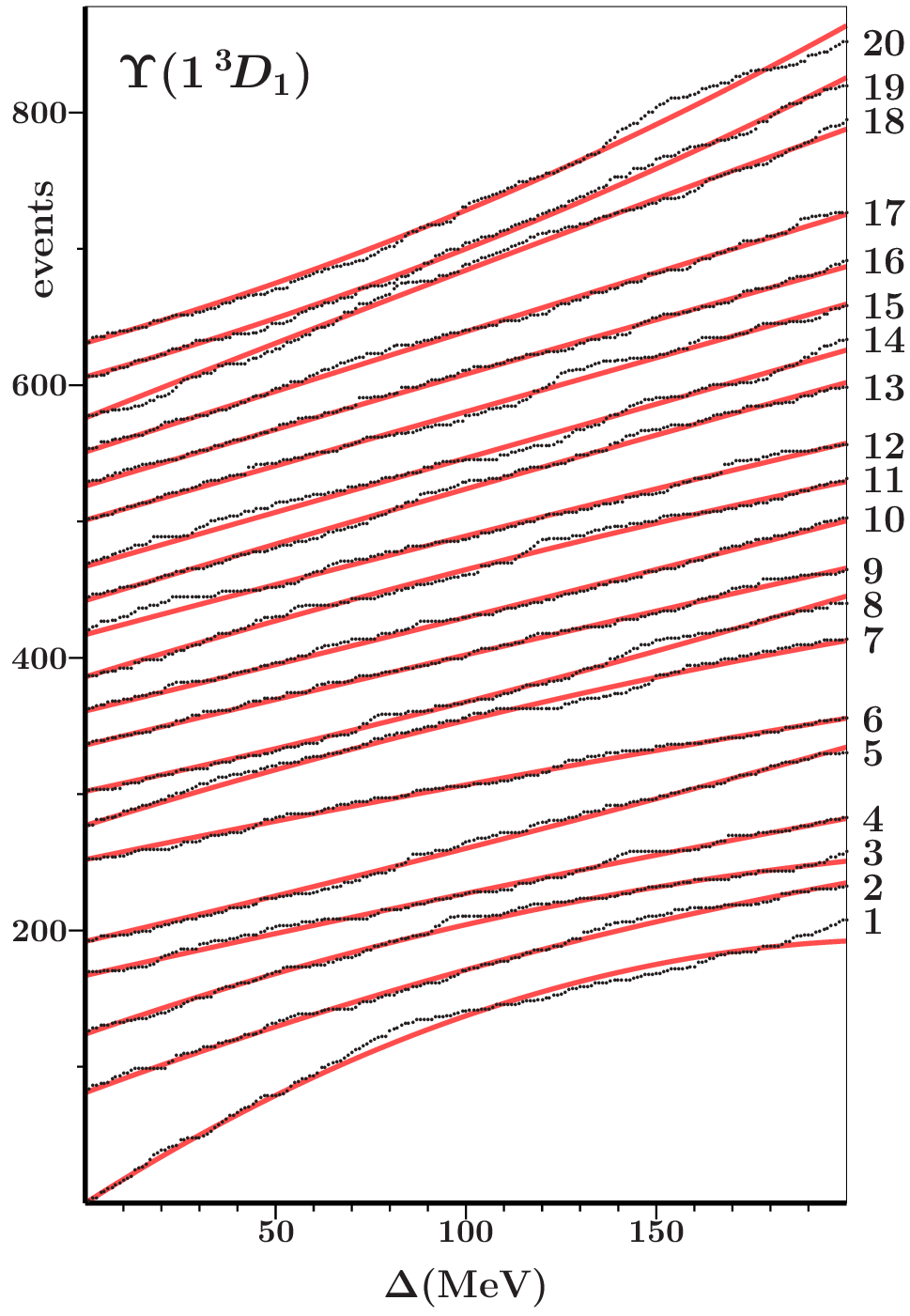}\\ [-10pt]
\end{tabular}
\end{center}
\caption{\small
Number of events
for the reaction $e^{+}e^{-}$ $\to$ $\pi^{+}\pi^{-}\Upsilon (1S)$
$\to$ $\pi^{+}\pi^{-}e^{+}e^{-}$,
scaled up by a constant, for increasing window openings
$\Delta$, from 1 MeV to 200 Mev in steps of 1 MeV,
in each of the 20 bins of the inset of Fig.~\ref{1Ddata}.
The corresponding bin numbers
are given on the right-hand vertical axis.
Each data set corresponding to a single bin
is fitted by a curve quadratic in $\Delta$ (solid lines)
that vanishes for $\Delta=0$.
The data are taken from Ref.~\cite{PRD78p112002}.
}
\label{all1Dbin}
\end{figure}

In Fig.~\ref{1Ddata} we have also indicated,
by error bars,
the standard deviation on the number of events in each bin.
For the data bins in the inset of Fig.~\ref{1Ddata},
we shall describe in detail how these values are determined.
To that end, we study how the number of events in each bin
grows when the window ($\Delta$) is increased from
0 to 200 MeV, in steps of 1 MeV.
The result is depicted in Fig.~\ref{all1Dbin}.

In order to clarify the relation between each set of data
in Fig.~\ref{all1Dbin} and the corresponding distributions shown
in Fig.~\ref{1Ddata},
we have numbered the bins in the inset of Fig.~\ref{1Ddata}
as well as the sets of data in Fig.~\ref{all1Dbin}.
Of course, each set of data in the latter figure
should start out at 0 events for $\Delta=0$ (window closed).
However, plotting the data this way
would result in a confusion of points and curves.
For clarity, we have thus increased the number of events
in Fig.~\ref{all1Dbin} by a constant, for each set of data.
Hence, for all data sets the number of events in reality
vanishes where the distributions meet the left-hand vertical
axis. So the annotations on the axis denote the scale, and
not the number of events corresponding to each data point.

Each of the data sets corresponding to a particular
interval in invariant $e^{+}e^{-}$ mass
is fitted with a quadratic curve,
where we assume that an infinite data set
would lead to a smoothly growing number of events
according as the window is further opened.
From the standard deviation of the actual data
with respect to smoothness,
we determine the size of the uncertainty
in the number of events in each bin,
also for the bins not shown in Fig.~\ref{all1Dbin}.
One may easily verify that the sizes of the error bars
shown in Fig.~\ref{1Ddata} are in agreement
with the deviations from smoothness in Fig.~\ref{all1Dbin}.

One observes in Fig.~\ref{all1Dbin}
that the number of events in bin no.~7
increases faster for small values of $\Delta$
than the number of events in the neighboring bins.
We believe this can be explained
by assuming that events from the narrow-width $\Upsilon(\ntdj{1}{1})$
fall well within the 10 MeV bin for not too large windows,
but start spilling to neighboring bins
when the window is further opened.
In the following, this issue will be dealt with more quantitatively.

\begin{figure}[htbp]
\begin{center}
\begin{tabular}{c}
\includegraphics[height=160pt]{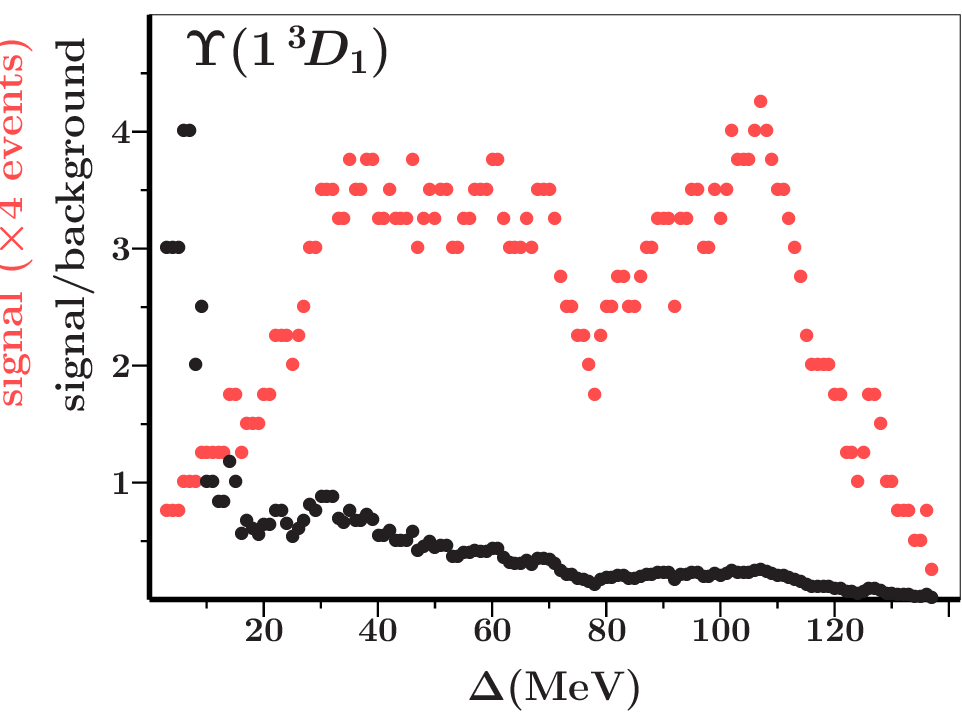}\\ [-10pt]
\end{tabular}
\end{center}
\caption{\small
Signal height (gray dots, pink in the online version)
and signal-to-background ratio (black)
for increasing window openings $\Delta$
from 1 MeV to 140 Mev in steps of 1 MeV,
for bin no.~7 of the inset of Fig.~\ref{1Ddata}.
Annotations on the vertical axis
correspond to the signal-to-background ratio;
the signal heights are 4 times larger.
}
\label{1Dsigback}
\end{figure}

In Fig.~\ref{1Dsigback} we study,
for the chosen signal bin
(number 7 in the inset of Fig.~\ref{1Ddata}),
the amount of events and also the signal-to-background ratio,
as a function of the window size $\Delta$.
For that purpose, we determine the background signal
by choosing the maximum of the number of events
in the neighboring bins
(numbers 6 and 8 in the inset of Fig.~\ref{1Ddata}).

We have also studied all other bins shown in the inset
of Fig.~\ref{1Ddata}.
None of these has a behavior for the signal height
and the signal-to-background ratio
that is even comparable to what we observe in Fig.~\ref{1Dsigback}
for bin no.~7.
For example, bin no.~12, at 10.15 GeV,
has a maximum signal height of 8 events, which vanishes
for windows larger than $\Delta =32$,
while bin no.~13, at 10.16 GeV, has no positive
signal for windows smaller than 80 MeV and, moreover,
a totally negligible signal-to-background ratio
for even wider windows.
From all this we may conclude that, in the invariant-mass interval
from 10.05 to 10.22 GeV, only one bin shows a very stable signal
for all possible windows, and a reasonable signal-to-background ratio
for windows up to about 70 MeV.

We assume that, for wider windows, possible events associated
with the supposed $\Upsilon(\ntdj{1}{1})$ signal start to spill over
to the neighboring bins.
As a consequence, the background increases, whereas
signal is lost.
Hence, we prefer to study the statistical relevance for
a smaller window, but still large enough for the signal height to be
already reasonable.
One may conclude from Fig.~\ref{1Dsigback} that taking
windows ranging from $\Delta =25$ MeV to $\Delta =35$ MeV is a good choice,
since for lower values the signal is relatively small,
whereas for higher ones the signal-to-background ratio decreases rapidly.
For the purpose of our data analysis, we have chosen $\Delta =30$ MeV,
as shown in Fig.~\ref{1Ddata}.

The relevant bin for the $\Upsilon(\ntdj{1}{1})$ contains $30\pm3.07$
events. For this bin, a global fit to the data (see Fig.~\ref{1Ddata})
gives a background of 18.0$\pm$2.53 events,
leaving 12.0$\pm$3.98 events for the signal.
Thus, the signal has a significance of 3.0$\sigma$,
which implies a strong indication of the $\Upsilon(\ntdj{1}{1})$
at 10098$\pm$5 MeV.

For the uncertainty in the mass of the $\Upsilon(\ntdj{1}{1})$,
we take half the bin width,
as we do not expect this state
to have a width larger than 1 MeV.
Consequently, the scattering of the data within the 10-MeV-wide bin
entirely stems from the method to collect data.

The difference between the aforementioned value of $10.113$ GeV
and the observed mass can be understood from
the predicted small mass shifts of ${}^{3\!}D_{1}$
bottomonium states due to open-bottom meson-loops \cite{PRD21p772}.

The CLEO Collaboration announced \cite{PRD70p032001}
the discovery of a $J^{PC}=2^{--}$ $b\bar{b}$ resonance
with a central mass of (10161.1$\pm$0.6$\pm$1.6) MeV.
Although CLEO reported
a significance of 10.2 standard deviations,
their observation is still today omitted from the Summary Tables
of the Particle Data Group \cite{JPG37p075021}.
Our identification of the $1^{--}$
$\Upsilon(\ntdj{1}{1})$ state, which comes out 63 MeV below the
$2^{--}$ $\Upsilon(\ntdj{1}{2})$,
makes CLEO's assignment of quantum numbers very plausible,
also in view of the quite different couplings of vector and tensor
states to open-bottom channels.

The observation of CLEO was recently confirmed by BABAR
\cite{ARXIV10040175}, obtaining a value of
$10164.5\pm 0.8_\xscrpt{stat}\pm 0.5_\xscrpt{syst}$ MeV
for the $\Upsilon(\ntdj{1}{2})$ mass.
However, in the same paper BABAR claimed
a mass of 10151 MeV for the $\Upsilon(\ntdj{1}{1})$,
which is close to the model prediction in
Ref.~\cite{PRD32p189}, but about 53 MeV above our value.
In the present study, as mentioned before,
we do not find any other relevant signal
in the invariant-mass interval from 10.05 to 10.22~GeV.

\begin{figure}[htbp]
\begin{center}
\begin{tabular}{c}
\includegraphics[height=330pt]{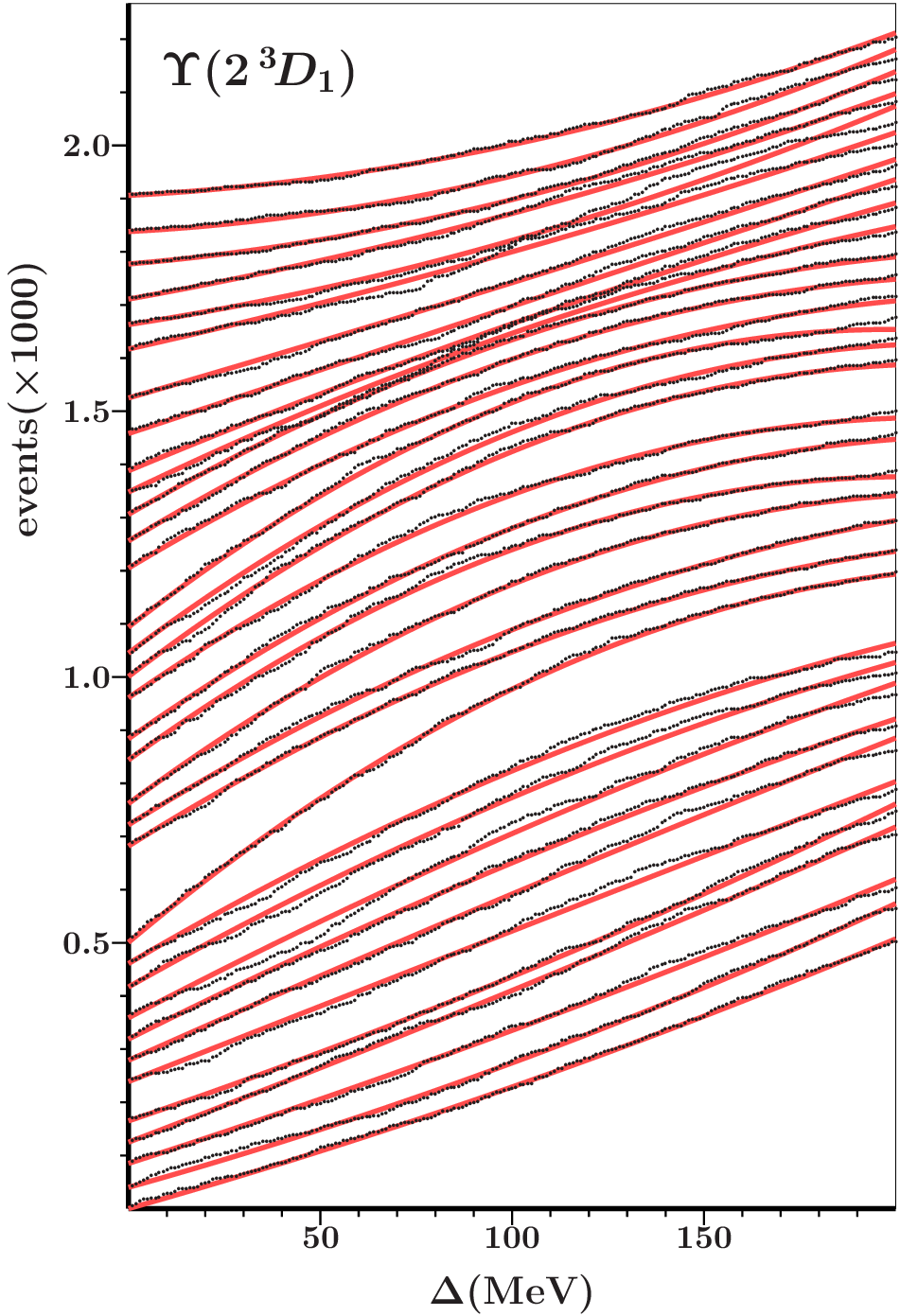}\\ [-10pt]
\end{tabular}
\end{center}
\caption{\small
Number of events
for the reaction $e^{+}e^{-}$ $\to$ $\pi^{+}\pi^{-}\Upsilon (1S)$
$\to$ $\pi^{+}\pi^{-}e^{+}e^{-}$,
scaled up by a constant, for increasing window openings
$\Delta$, from 1 MeV to 200 Mev in steps of 1 MeV,
in each of the 34 bins of Fig.~\ref{2Ddata}.
Each data set corresponding to a single bin
is fitted by a curve quadratic in $\Delta$ (solid lines)
that vanishes for $\Delta=0$.
The data are taken from Ref.~\cite{PRD78p112002}.
}
\label{all2Dbin}
\end{figure}

\begin{figure}[htbp]
\begin{center}
\begin{tabular}{c}
\includegraphics[height=160pt]{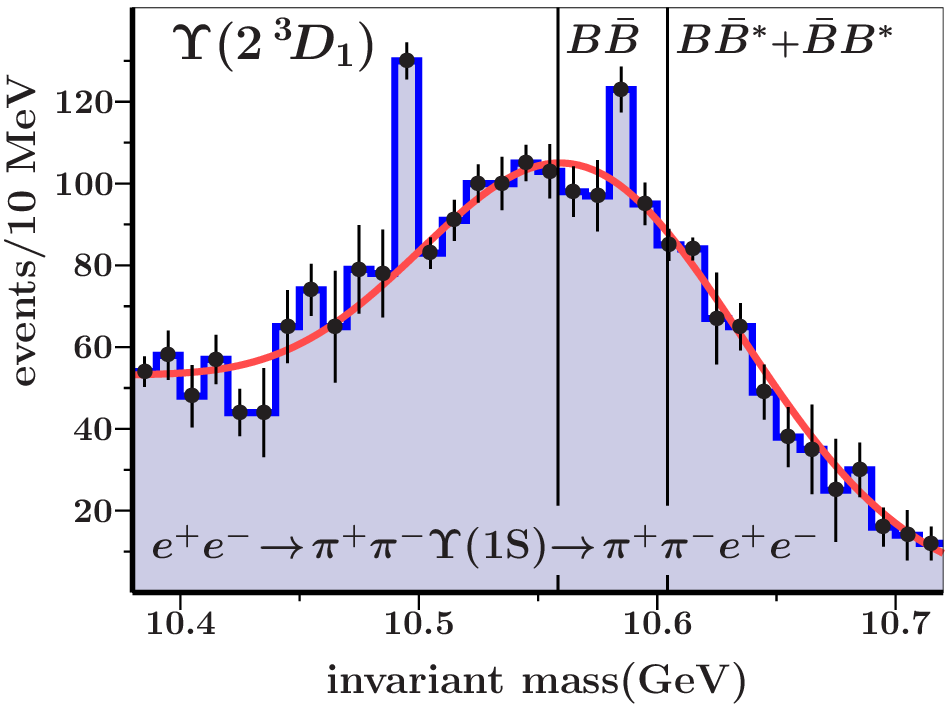}\\ [-10pt]
\end{tabular}
\end{center}
\caption{\small
Invariant-mass distribution for 2311 events in
$e^{+}e^{-}$ $\to$ $\pi^{+}\pi^{-}\Upsilon (1S)$
$\to$ $\pi^{+}\pi^{-}e^{+}e^{-}$,
for window and bin sizes of $\Delta =22$
and 10 MeV, respectively.
Data are taken from Ref.~\cite{PRD78p112002}.
Our fit to the background is shown by a solid curve.
The $\Upsilon(\ntdj{2}{1})$ signal has a significance of 10.7$\sigma$.
For completeness, we have also indicated
the $B\bar{B}$ and $B\bar{B}^{\ast}+\bar{B}B^{\ast}$ thresholds.
}
\label{2Ddata}
\end{figure}

For the signal stemming from the $\Upsilon(\ntdj{2}{1})$,
we follow precisely the same strategy as discussed above
for the $\Upsilon(\ntdj{1}{1})$.
In Fig.~\ref{all2Dbin} we display how the number of events
develops when the window ($\Delta$) is increased from
0 to 200 MeV, in steps of 1 MeV,
for all 10-MeV-wide bins shown in Fig.~\ref{2Ddata}.
The sets of data in Fig.~\ref{all2Dbin},
from the bottom to the top curve,
correspond to the bins in Fig.~\ref{2Ddata}
from left to right.
For example, data set no.~12 counted from below in
Fig.~\ref{all2Dbin}, which starts out at the annotation
for 500 events on the left-hand vertical axis,
corresponds to bin no.~12 counted from the left in
Fig.~\ref{2Ddata}. This is the bin that contains more events
than any other shown in Fig.~\ref{2Ddata}.
We observe for the latter set of data
a substantially faster rise in the number of events
for increasing values of the window size $\Delta$,
up to about $\Delta =120$,
than for the data sets associated with the neighboring bins.
The most obvious explanation for this behavior
is the presence of an enhancement in bin no.~12,
for which the most likely candidate is the
$\Upsilon(\ntdj{2}{1})$.

\begin{figure}[htbp]
\begin{center}
\begin{tabular}{c}
\includegraphics[height=160pt]{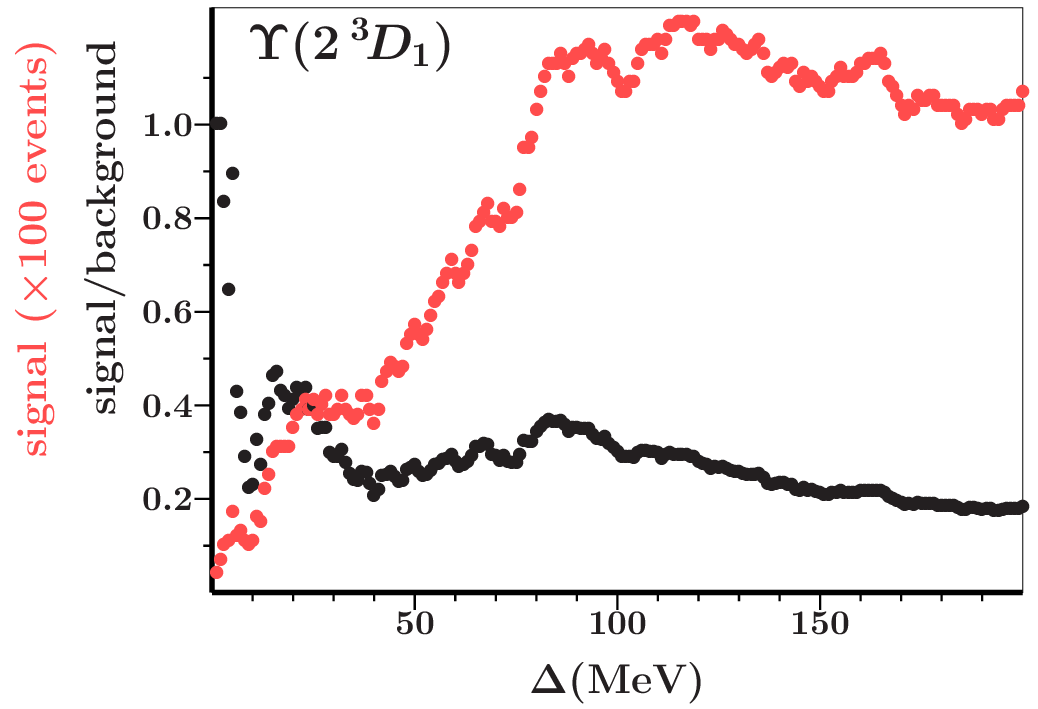}\\ [-10pt]
\end{tabular}
\end{center}
\caption{\small
Signal height (gray dots, pink in the online version)
and signal-to-background ratio (black)
for increasing window openings $\Delta$,
from 1 MeV to 200 Mev in steps of 1 MeV,
for the bin containing the $\Upsilon(\ntdj{2}{1})$ signal
of Fig.~\ref{2Ddata}.
Annotations on the vertical axis
correspond to the signal-to-background ratio;
the signal heights are 100 times larger.
}
\label{2Dsigback}
\end{figure}

\begin{figure}[htbp]
\begin{center}
\begin{tabular}{c}
\includegraphics[height=160pt]{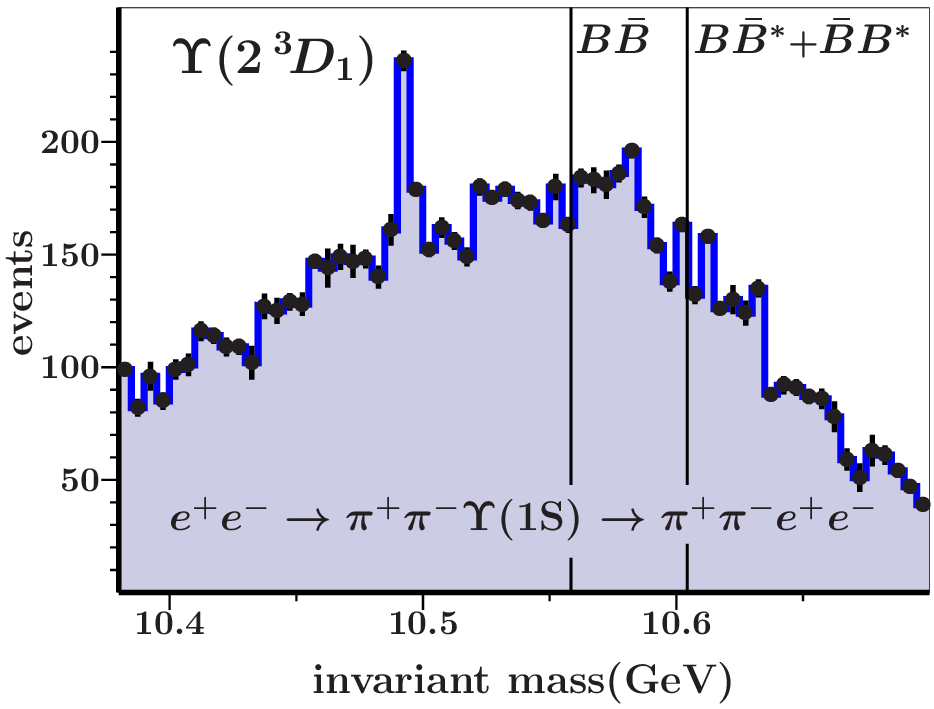}\\ [-10pt]
\end{tabular}
\end{center}
\caption{\small
Invariant-mass distribution for 8337 events in
$e^{+}e^{-}$ $\to$ $\pi^{+}\pi^{-}\Upsilon (1S)$
$\to$ $\pi^{+}\pi^{-}e^{+}e^{-}$
for window and bin sizes of $\Delta =82$
and 5 MeV, respectively.
Data are taken from Ref.~\cite{PRD78p112002}.
For completeness, we have also indicated
the $B\bar{B}$ and $B\bar{B}^{\ast}+\bar{B}B^{\ast}$ thresholds.
}
\label{2D5bin}
\end{figure}

In Fig.~\ref{2Dsigback} we depict
the signal and the signal-to-background ratio
for the relevant bin.
From this result we conclude that a window of $\Delta =22$ MeV
is appropriate for analysis.
However, we admit that an equally good analysis is certainly possible
for a window of $\Delta =82$ MeV.
We show the data for the latter window in Fig.~\ref{2D5bin},
using a data binning of 5 MeV,
but shall not further analyse this case, for the same reason as we
have discussed for the $\Upsilon(\ntdj{1}{1})$.
Namely, for larger windows we suspect data to spill over
over to neighboring bins.

The data for the $\Upsilon(\ntdj{2}{1})$
are depicted in Fig.~\ref{2Ddata}.
The relevant bin contains 130$\pm$4.51 events,
while from a fit to the global data
(solid curve in Fig.~\ref{2Ddata})
we find a background signal of 81$\pm$0.81 events.
For the signal we thus obtain 49$\pm$4.58 events,
and so a significance of 10.7$\sigma$.

Apparently, the $^{3\!}D_1$ $b\bar{b}$ states couple quite weakly to the
transitions under study here, as can been seen from the large
$\ntsj{2}{1}$ and $\ntsj{3}{1}$ peaks in Fig.~\ref{1Ddata}, to be contrasted
with the more modest $\ntdj{2}{1}$ (Figs.~\ref{2Ddata},\ref{2D5bin}) and
especially $\ntdj{1}{1}$ (Fig.~\ref{1Ddata}) structures. This observation
is in line with expectations from theory, since $D$-wave quarkonium states
are suppressed in $e^+e^-$ annihilation, due to the small wave function at
short distances. However, the much stronger signal for the $\ntdj{2}{1}$ state
as compared to the $\ntdj{1}{1}$ seems puzzling at first sight, also in view
of the $\ntsj{2}{1}$ peak, which is higher than the $\ntsj{3}{1}$. A possible
explanation is offered by coupled channels. Namely, even $b\bar{b}$ systems
below the lowest open-bottom threshold ($B\bar{B}$) couple to virtual meson
pairs, both the $\tsj{1}$ and $\tdj{1}$ states. This coupling inevitably gives
rise to $\tsj{1}$-$\tdj{1}$ mixing. Now, the farther the $b\bar{b}$ system
lies below the $B\bar{B}$ threshold, the more virtual will the meson pairs be,
and so the smaller the $\tsj{1}$-$\tdj{1}$ mixing. Therefore, the
$\Upsilon(\ntdj{2}{1})$, which lies only 64 MeV below the $B\bar{B}$
threshold, will have a significantly larger $\tsj{1}$ admixture than the
almost 400 MeV lighter $\Upsilon(\ntdj{1}{1})$, and so its effective coupling
to $e^+e^-$ will be considerably larger.

Nevertheless, the smaller $\tdj{1}$ signals
may explain why they have not been discovered
over the past three decades of $e^{+}e^{-}$ accelerator physics.
Moreover, these states are probably also quite narrow.
This complicates event selection,
since on the one hand we prefer a narrow bin size for high resolution,
but on the other hand we would also like to have sufficient statistics,
which can be achieved by choosing large windows
for the invariant masses of the final-state $e^{+}e^{-}$ pair
around the $\Upsilon (1S)$ mass.
However, too large windows inevitably produce
several non-candidates and a very noisy background,
as may be observed by comparing the distributions
shown in Figs.~\ref{2Ddata} and \ref{2D5bin}.
The choices we have made in the present analysis
reflect a compromise between the two criteria.

The here reported $\Upsilon(\ntdj{1}{1})$ and $\Upsilon(\ntdj{2}{1})$
masses lend further support to the level scheme of
the harmonic-oscillator (HO) approximation to
the resonance-spectrum expansion (HORSE)
(see Ref.~\cite{ARXIV09100967} and references therein).
In particular, the observed spacing of about 380 MeV between the
$D$ states \cite{ARXIV10091778} makes a $\ntdj{3}{1}$ assignment for
the $\Upsilon (10860)$ more plausible than $\ntsj{5}{1}$.

To be more precise, the pure HO, without accounting for meson loops,
gives a mass of $2\times 4.724+0.19\times(4+2+1.5)=10.873$ GeV
\cite{PRD27p1527},
BABAR obtained 10.876$\pm$0.002 GeV \cite{PRL102p012001},
while in Ref.~\cite{ARXIV09100967} we found 10.867 GeV
for the $\Upsilon(10860)$ mass.
Moreover, this result supports
the conclusion of our analysis in Ref.~\cite{ARXIV09100967}
that the $\Upsilon (10580)$ is not the $\Upsilon (4S)$ resonance,
but rather a $B\bar{B}$ threshold enhancement \cite{PRD80p074001}.
For the $\Upsilon (4S)$ resonance, which in the HORSE is expected
some 100--200 MeV below the $\Upsilon(3D)$, due to the effect of the meson
loops, we found in Ref.~\cite{ARXIV09100967} a central mass of 10.735 GeV,
by analyzing BABAR data published in Ref.~\cite{PRL102p012001}. Finally,
the conspicuity of the $\ntdj{2}{1}$ signal, when compared to the much
feebler $\ntdj{1}{1}$, is a compelling indication of virtual coupled-channel
effects, as explained above. This represents an additional endorsement of the
HORSE and its nonperturbative description of decay, including open as well
as closed meson-meson channels.

In conclusion, we have observed excellent candidates for the
$\Upsilon(\ntdj{1}{1})$ and $\Upsilon(\ntdj{2}{1})$ states,
with central masses of 10098$\pm$5 MeV and 10495$\pm$5 MeV, respectively.
Their observation, if confirmed independently, strongly supports the
resonance level scheme of the harmonic-oscillator approximation to
the resonance-spectrum expansion for quarkonia.

We are grateful for the precise measurements
of the BaBar Collaboration,
which made the present analysis possible.
We wish to thank Dr. Chengping Shen for useful remarks.
This work was supported in part by the {\it Funda\c{c}\~{a}o para a
Ci\^{e}ncia e a Tecnologia} \/of the {\it Minist\'{e}rio da Ci\^{e}ncia,
Tecnologia e Ensino Superior} \/of Portugal, under contract
CERN/\-FP/\-109307/\-2009.

\newcommand{\pubprt}[4]{#1 {\bf #2}, #3 (#4)}
\newcommand{\ertbid}[4]{[Erratum-ibid.~#1 {\bf #2}, #3 (#4)]}
\def\JPG{J.\ Phys.\ G}
\def\PRD{Phys.\ Rev.\ D}
\def\PRL{Phys.\ Rev.\ Lett.}

\end{document}